\documentclass[]{interact}
\usepackage[table]{xcolor} 
\usepackage{epstopdf}
\usepackage[caption=false]{subfig}
\usepackage[numbers,sort&compress]{natbib}
\bibpunct[, ]{[}{]}{,}{n}{,}{,}

\theoremstyle{plain}
 \usepackage{url} 
\begin{document}

\articletype{}

\title{Quantifying Uncertainty: All We Need is the Bootstrap?}

\author{
\name{Ur\v{s}a Zrim\v{s}ek\textsuperscript{a} and Erik \v{S}trumbelj\thanks{CONTACT E. \v{S}trumbelj. Email: erik.strumbelj@fri.uni-lj.si}\textsuperscript{a}}
\affil{\textsuperscript{a}Faculty of Computer and Information Science, University of Ljubljana, Ve\v{c}na pot 113, 1000 Ljubljana, Slovenia}
}

\maketitle

\begin{abstract}
\{A critical literature review and comprehensive simulation study is used to show that (a) non-parametric bootstrap is a viable alternative to commonly taught and used methods in basic estimation tasks (mean, variance, quartiles, correlation) and (b), contrary to recommendations in most related work, double bootstrap performs better than BCa.\} Quantifying uncertainty through standard errors, confidence intervals, hypothesis tests, and related measures is a fundamental aspect of statistical practice. However, these techniques involve a variety of methods, mathematical formulas, and underlying concepts, which can be complex. Could the non-parametric bootstrap, known for its simplicity and general applicability, serve as a universal alternative? This paper addresses this question through a review of the existing literature and a simulation analysis of one- and two-sided confidence intervals across varying sample sizes, confidence levels, data-generating processes, and statistical functionals. Results show that the double bootstrap consistently performs best and is a promising alternative to traditional methods used for common statistical tasks. These results suggest that the bootstrap, particularly the double bootstrap, could simplify statistical education and practice without compromising effectiveness.
\end{abstract}

\begin{keywords}
Statistics; inference; standard errors; confidence intervals; simulation study
\end{keywords}

\section{Introduction}

\setlength{\tabcolsep}{3pt}

University curricula in fields such as the social sciences, medicine, and life sciences, which heavily rely on statistical methodology, typically include only one or two applied statistics courses. However, it is practitioners in these fields, rather than professional statisticians, who perform the majority of statistical analyses. The mismatch between the level of training provided and the practical demand for statistical analysis often results in an over-reliance on rote memorization and formulaic application of methods, contributing to challenges such as the replication crisis in science. This underscores the importance of exploring ways to simplify current statistical practices. Simplification could not only enhance comprehension and reduce errors but also create opportunities to incorporate other aspects of statistical methodology into curricula.

This paper focuses on the quantification of uncertainty. Standard errors, confidence intervals, and hypothesis tests are integral components of statistical practice, yet they typically involve advanced concepts—such as test statistics and sampling distributions—and encompass a wide array of methods. Among these, one method stands out: the bootstrap. The bootstrap offers several advantages over traditional techniques for quantifying uncertainty. It is conceptually straightforward, reinforces the fundamental role of sampling in statistics, allows direct interaction with estimates and their distributions, and can be applied to a wide range of practical tasks without requiring the mastery of new concepts or complex mathematical formulas. These characteristics give the bootstrap significant pedagogical value \citep{Hesterberg2015}, positioning it as a strong candidate for a one-size-fits-all approach to quantifying uncertainty, particularly for practitioners with limited statistical training.

Historically, the use of the bootstrap was constrained by computational limitations. With advancements in computing power, this is no longer a concern, yet introductory applied statistics textbooks and courses rarely place the bootstrap at the forefront. This is largely due to the inertia of established practices, which are supported by more extensive instructional resources and software tools—resources that bootstrapping has historically lacked \citep{Hesterberg2015}, and to some extent still does. Moreover, there is currently insufficient empirical evidence to convincingly establish the bootstrap as a comprehensive, one-size-fits-all solution for quantifying uncertainty.

\subsection{Related work}\label{sec:related}

Theoretical cases where bootstrap fails are pathological with limited implications for practice (see \cite{carpenter2000bootstrap} and \cite[Ch.2.6]{davison1997bootstrap}). Large sample properties of common bootstrap approaches have also been established (see \cite{carpenter2000bootstrap} for a summary). Unfortunately, large sample theory is not always a reliable predictor of finite sample performance, so empirical work is required.

Tables \ref{tbl:rw1} and \ref{tbl:rw2} summarize empirical studies on bootstrap methods. Early research primarily focused on the Pearson correlation and the sample mean, with subsequent work extending to other functionals, particularly quantiles and regression parameters. In this review, the focus is on non-parametric bootstrap techniques and methods that are widely taught and applied in practice, referred to hereafter as \textbf{baseline} methods. These include, for example, Fisher confidence intervals (CIs) for the Pearson correlation and t-intervals for the sample mean.

While most, though not all, related studies include a baseline comparison, there are only three instances—aside from an early study where the 1st-order accurate percentile bootstrap (\textbf{PB}) was the best-performing method \citep{rasmussen1987estimating}—in which the baseline outperforms the top-performing bootstrap approach. For example, \cite{Dorfman1990} demonstrated that Fieller and Taylor series-based confidence intervals for elasticities and flexibilities marginally outperformed the bias-corrected and accelerated (\textbf{BCa}), bias-corrected (\textbf{BC}), and PB methods, though this result was observed with relatively small sample sizes ($n = 13, 15$). Similarly, \cite{puth2015} showed that for Pearson correlation under a bivariate normal distribution with varying correlation coefficients, Fisher intervals achieved coverage close to the nominal 95\%, while BCa intervals yielded coverage rates around 93\% and 94\% for sample sizes of 20 and 10, respectively. Furthermore, \cite{ialongo2019confidence} found that for the log-normal distribution, baseline methods attained nominal coverage with smaller sample sizes than BCa, though this was contingent on assuming log-normality.

The most common recommendation propagated through literature is to use BCa, based mostly on theoretical results. Empirical results also often recommend BCa, with the exception of the mean, where studentized bootstrap (also bootstrap-t, \textbf{B-t}) is better. However, some studies suggest that BCa does not perform well with small sample sizes \citep{carpenter2000bootstrap,chernick2009revisiting,owen1988empirical}. Double or iterated (also calibrated) bootstrap (\textbf{DB}) appears in only 7 studies. When it does, it performs as well as or better than other methods. 

In summary, much of the related research is confined to a single functional, a single data-generating process (DGP), and/or a single confidence level. Additionally, the most commonly used methods are not always included as baselines for comparison. DB also warrants greater attention. While the findings are promising, it remains challenging to fully assess the practical implications of relying exclusively on bootstrap methods.

\begin{table}[!htbp]
\centering
\tbl{A summary of simulation studies in related work (1981-1999). Ordered by year of publication.}{
\begin{tabular}{
  p{0.03\linewidth} 
  p{0.10\linewidth} 
  p{0.07\linewidth} 
  p{0.035\linewidth} 
  p{0.10\linewidth} 
  p{0.13\linewidth} 
  p{0.15\linewidth}
  p{0.3\linewidth}}\toprule
  \rowcolor{gray!50}  
  Ref. & DGP & n & n$_\text{rep}$ & Functional & Evaluating & Methods & Summary of results \\ \midrule 
    \rowcolor{gray!10}   \cite{efron1981nonparametric} & normal & 14 & 200 & mean & SE & PB, SB (B = 128, 512) & SB better than PB \\
  \cite{schenker1985qualms} & normal & 20-100 & 1600 & variance & 90\% CI & BC, PB (B = 1000) & bootstrap coverage below nominal \\   
      \rowcolor{gray!10} \cite{efron1986bootstrap} & normal, exponential & 15 & 200 & trimmed mean & SE & PB (B = 200) & PB better than jackknife \\  
  \cite{rasmussen1987estimating}  & normal, non-normal & 5-60 & 1000 & mean & 95\%, 99\% CI & baseline; PB (B = 500) & baseline better than PB \\
      \rowcolor{gray!10} \cite{Silverman1987} & normal & 14-100 & 200 & mean & SE & PB, SB (B = 200) & SB better than PB \\
  \cite{diciccio1988review}  & exponential & 5 & 1 & corr & 95\% CI & BC, BCa, PB (B = ?) & BCa good; PB and BC poor\\   
  & bivariate non-normal & 5 & 1 & ratio & 95\% CI & BC, BCa, PB (B = ?) & BCa good; PB and BC poor\\    
  & normal & 8 & 1 & mean & 95\% CI & B-t, BC, PB (B = ?) & BC and PB perform similarly, B-t is worse \\
     \rowcolor{gray!10}  \cite{efron1988bootstrap} & normal & 100 & 100 & mean & several CI & baseline; PB, SB (B = 2000) & PB similar to baseline; SB better than PB \\ 
  \cite{owen1988empirical} & chi-squared & 20 & 1000 & corr & 90\% CI & baseline; hybrid, B-t, BC, BCa, PB (B = 1000) & hybrid, B-t, and PB perform similarly; BC and BCa have lower than nominal coverage\\    
      \rowcolor{gray!10} \cite{Dorfman1990} & (see paper) & 13, 15 & 500 & regression & 90\% CI & baseline; BC, BCa, PB (B = 500) & baseline better than bootstrap; bootstrap methods similar to each other\\     
  \cite{young1990bootstrap} & normal, exponential & 10, 20 & 1000 & corr & several CI & PB (B = 50000) & PB is biased \\  
     \rowcolor{gray!10}  \cite{owen1992empirical} & 7 distributions (see paper) & 3-20 & 1000  & corr & 95\% CI & baseline; B-t, BCa, PB (B = 1000) &  B-t performs well \\  
  \cite{mammen1992}  & chi-squared, mixture of normal & 20 & 10000  & corr & several CI & baseline; B-t, PB (B = 1000) & bootstrap similar to baseline; B-t is best\\ 
      \rowcolor{gray!10} \cite{shi1992accurate} & normal, Poisson, t, Weibull & 20 & 500 & corr & 90\% CI & baseline; B-t, BCa, DB, PB (B = 500) & DB and B-t perform best \\  
  \cite{Stanny1993} & normal, exponential & 5-25 & 1000 & corr & 95\% CI & baseline; DB, PB (B = 1000) & DB as good or better than baseline, except for $n = 5$; PB is worst \\
      \rowcolor{gray!10} \cite{Lee1994} & normal, exponential, beta, gamma, t & 10, 100 & 500  & median & SE & PB, SB (B = 200) & smoothing improves performance \\     
  \cite{diciccio1996bootstrap} & normal & 20  & 1 & mean & 90\% CI & baseline; ABC, BCa, B-t (B = 2000) & bootstrap better than baseline \\ 
      \rowcolor{gray!10} \cite{Lee1996} & normal, folded normal, exponential, log-normal & 15, 30 & 1600  & corr & 90\% CI  &  ABC, DB, DB-ABC, PB (B = 1000) & DB better than ABC and calibrated ABC \\   
  \cite{Letson1998} & (see paper) & 13, 15 & 500 & regression & 90\% CI & DB, PB (B = 1999) & DB better than PB \\ \bottomrule    
\end{tabular}}\tabnote{\textbf{n} = sample size or range, if more than two; \textbf{ n$_\text{rep}$ }= number of Monte Carlo replications; \textbf{B} = number of bootstrap replications; \textbf{?} indicates information that could not be discerned from the paper} \label{tbl:rw1}
\end{table}

\begin{table}[!htbp]
\centering
\tbl{A summary of simulation studies in related work (2000-2023). Ordered by year of publication.}{
\begin{tabular}{
  p{0.03\linewidth} 
  p{0.10\linewidth} 
  p{0.07\linewidth} 
  p{0.035\linewidth} 
  p{0.10\linewidth} 
  p{0.13\linewidth} 
  p{0.15\linewidth}
  p{0.3\linewidth}}\toprule
  \rowcolor{gray!50}  
  Ref. & DGP & n & n$_\text{rep}$ & Functional & Evaluating & Methods & Summary of results \\\midrule    
     \rowcolor{gray!10} \cite{carpenter2000bootstrap} &  inverse exponential & 20 & 10000 & corr & 99\% CI &  baseline; B-t, BC, BCa, PB (B = 4999) & baseline, B-t, and BCa perform similarly; BCa poor for small n\\       
  \cite{efron2003second} & normal & 15 & 1 & mean &  90\% CI & baseline; ABC & bootstrap better than baseline \\    
     \rowcolor{gray!10}  \cite{zhou2005nonparametric} & log-normal, gamma & 10-50 & 10000 & diff. in means & 95\% CI & baseline; B-t, BCa (B = 1000) & B-t performs well and better than BCa\\  
  \cite{arasan2008alternative} & bivariate normal (censored) & 25-400 & 2000 & (see paper) & 90\%, 95\% CI & B-t, BCa, PB (B = 520) & jackknife performs best; PB performs worst \\   
     \rowcolor{gray!10}  \cite{chernick2009revisiting} & normal, log-normal, gamma, t, uniform & 10-3600 & 1000-64000 & variance & several CI & baseline; ABC, BC, BCa, PB (B = 1000, 16000) & 2nd order accurate methods can converge slowly and perform worse than PB\\ 
  \cite{cheung2009comparison} & (see paper) & 50-500 & 5000    & regression & 95\% CI & baseline; BC, PB  (B = 2000) & BC and PB perform well \\  
     \rowcolor{gray!10}  \cite{hall2010bootstrap} & exponential, Pareto & 10-25 & 2000 & extrema & 80\%, 90\% CI & baseline; PB, DB (B = 699) & DB is best \\
  \cite{jones2013computing} & (see paper) & 50-500 & 5000 & regression & 95\% CI & baseline; BCa, PB (B = 10000) & BCa and PB perform well \\  
     \rowcolor{gray!10}  \cite{karavarsamis2013comparison} & (see paper) & 10-50 & 1000 & capture-recapture & 95\% CI & baseline; B-t, PB (B = 100, 250) & B-t performs best \\ 
  \cite{arasan2014double} & log-logistic with censored data & 25-50 & 1000 & (see paper) & 90\%, 95\% CI & baseline; B-t, DB, DB-t, PB (B = ?) & PB poor; other bootstrap methods good and similar to each other \\         
     \rowcolor{gray!10}  \cite{padilla2014bootstrapped} & normal, uniform, triangular, beta, Laplace, Pareto & 50?-300? & 1000 & mean & 95\% CI & BCa, PB (B = 2000) & BCa slightly better than PB; good coverage except on Pareto\\   
  \cite{Hesterberg2015} &  normal, exponential & 5-4000 & 10000 & corr & 95\% CI & baseline; B-t, PB, reverse PB (B = 10000) & bootstrapped CI narrow, especially for small n; PB worse than baseline for small n; reverse PB poor; B-t best\\   
     \rowcolor{gray!10}  \cite{loh2015inferential} & log-logistic with censored data & 25-60 & 1000 & (see paper) & 90\%, 95\% CI & baseline, JK; DB, PB  (B = 1000) & DB best; PB poor \\     
  \cite{puth2015} & normal & 10-100 & 10000 & mean & 95\% CI & baseline; B-n, B-t, BC, BCa, PB  (B = 10000) & baseline better than bootstrap on small n; BCa performs well \\
     \rowcolor{gray!10}  \cite{flowers2018comparison} & (see paper) & ? & 1000 & quantiles & 95\% CI & modified B-n, BC, BCa, PB (B = 1999) & bootstrap methods perform similarly \\ 
  \cite{Ialongo2019} & log-normal & 20-120 & 100 & quantiles & 95\% CI & baseline; BCa, PB  (B = 10000) & baseline better than bootstrap for small n \\     
     \rowcolor{gray!10}  \cite{DelosReyes2023} & 26 pairs & 20-600 & 1000 & corr & 95\% CI & baseline; BCa, PB (B = 2000) & BCa better than PB on small n; PB better than BCa for large n; poor coverage for uniform-chi-squared \\   \bottomrule    
\end{tabular}}\tabnote{\textbf{n} = sample size or range, if more than two; \textbf{ n$_\text{rep}$ }= number of Monte Carlo replications; \textbf{B} = number of bootstrap replications; \textbf{?} indicates information that could not be discerned from the paper} \label{tbl:rw2}
\end{table}

\section{Simulation Study}\label{ch:experiment}

The experiment runs across each possible combination of sample size $n$ from \{4, 8, 16, 32, 64, 128, 256\}, intervals $(-\infty, \alpha)$ with endpoints $\alpha$ from \{0.025, 0.05, 0.25, 0.75, 0.95, 0.975\}, statistical functional from \{mean, median, standard deviation, $5^{th}$ and $95^{th}$ percentile, and Pearson correlation\}, and data generating process from

\begin{itemize}
    \item normal with $\mu = 0$ and $\sigma = 1$,
    \item exponential with $\lambda = 1$, 
    \item uniform from 0 to 1,
    \item beta with $\alpha = 10$ and $\beta = 2$,
    \item log-normal with $\mu = 0$ and $\sigma = 1$,
    \item Laplace with $\mu = 0$ and $b = 1$, 
    \item Bernoulli with $p = 0.5$,
    \item Bernoulli with $p = 0.9$, and
    \item bivariate normal with $\mu = 
            \begin{bmatrix}
            1 \\
            1
            \end{bmatrix}$ 
            and $\Sigma = 
            \begin{bmatrix}
            2 & 0.5\\
            0.5 & 1
            \end{bmatrix}$. 
\end{itemize}

\noindent Note that the bivariate normal and Pearson correlation appear only in combination with each other and the two Bernoulli distributions appear only in combination with the mean.

The experimental design was informed by a combination of related simulation studies, theoretical considerations, and computational constraints. The selected range of sample sizes is consistent with those commonly used in similar studies. However, sample sizes of 4 and 8, which are rarely examined, are included. Although such small samples are uncommon in most statistical practice, they may hold practical relevance in fields where data is limited (for example, gene expression studies). For sample sizes of 256 and above, all methods should perform well from a practical standpoint, though these larger sample sizes can also become computationally prohibitive.

The selected range of endpoints includes the 5\% and 95\% confidence levels and enables the construction of symmetric two-sided confidence intervals at 95\% and 90\% levels by combining the 0.025 and 0.975, or 0.05 and 0.95 endpoints. These levels are the most commonly used in practice and almost all simulation studies. However, this choice of endpoints also permits the exploration of other confidence levels. It is important to note, from an evaluation standpoint, that considering one-sided coverage is crucial, as relying solely on two-sided coverage can be misleading (see Section \ref{sec:measuring} for details).

The selected statistical functionals include the most commonly used functionals (mean, median, standard deviation, and correlation) and two extreme percentiles, which are less widely used in practice but are known to pose challenges in quantifying uncertainty. A notable omission is the inclusion of model coefficients, such as those from the family of generalized linear mixed-effects models. However, for many of these models, and most more complex models, the bootstrap is not only a viable alternative but the only option for quantifying uncertainty.

Every experiment is replicated $n_\text{rep} = 10000$ times, to limit coverage standard error to $0.005$ in the worst case, and $B = \{10, 100, 1000\}$ bootstrap replications. More bootstrap replications is better, so $B$ is not considered as a dimension of the experiment. Results are only reported for $B = 1000$, but $B = 100$ would result in the same main conclusions. For $B = 10$ the performance of bootstrap methods is noticeably worse. Recommendations \cite{davison1997bootstrap,chernick2009revisiting,efron1994introduction} and choices of $B$ in simulation studies also suggest that $B = 1000$ is sufficient.

\subsection{Methods}

CIs produced by the methods that are most commonly used in practice for that statistical functional are included as a baseline for comparison. For the \textit{mean}, the t-based CIs from the commonly used t-test (\textbf{t-test}). For the mean of the two Bernoulli distributions, the Clopper-Pearson (\textbf{c-p})~\citep{Clopper1934} and Agresti-Coull (\textbf{a-c})~\citep{Agresti1998} intervals.  For the \textit{median}, CIs from the \textit{Wilcoxon signed rank test}  (\textbf{wilcoxon})~\citep{bauer1972constructing}. For \textit{standard deviation}, \textit{chi-squared} CIs (\textbf{chi-sq})~\citep{snedecor1989statistical}. For Pearson correlation Fisher CIs (\textbf{fisher}) \citep{bonett2000sample}. For quantiles, parametric CIs based on normal assumption (\textbf{q-par}), non-parametric CI (\textbf{q-nonpar}) (see \citep{ialongo2019confidence} for both), and the Maritz-Jarrett method (\textbf{m-j}) \citep{maritz1978note}. 

The bootstrap procedure can be divided into two primary steps. First, \emph{bootstrap sampling} to generate the bootstrap distribution, which serves as an approximation of the sampling distribution of the functional of interest. And second, applying a \emph{bootstrap method} to construct a confidence interval.

Bootstrap sampling can be further categorized into parametric and non-parametric approaches. The parametric bootstrap assumes a specific distribution for the underlying population, $F$, and estimates the associated parameters from the observed data, $X = (x_1, x_2, \dots, x_n)$. In contrast, the non-parametric bootstrap infers properties of $F$ by resampling directly from $X$ without imposing distributional assumptions. Given that there are $n^n$ possible samples from resampling, the process is typically restricted to $B$ independent bootstrap samples to maintain computational feasibility. This yields $B$ bootstrap samples $X^*_1, X^*_2, \dots, X^*_{B}$ and the bootstrap distribution of the parameter $\hat{\theta}^* = (\hat{\theta}^*_1, \hat{\theta}^*_2, \dots, \hat{\theta}^*_B)$.

The assumptions underlying parametric bootstrap methods restrict their applicability or require additional user input, making them less suitable as a universal approach. Therefore, the experiments focus exclusively on non-parametric bootstrap methods. Below is a brief overview of the bootstrap methods used for CI construction, along with references for further details. The source code is also available (see Section \ref{ch:results}).

\subsubsection{Percentile Bootstrap (\textbf{PB})}

The percentile method is the original method proposed by Efron in~\cite{efron1979another} (for details, also see~\cite[chap.~13]{efron1994introduction}). Multiple improvements have been made since, but percentile remains one of the most popular bootstrap methods. The percentile CI for confidence level $\alpha$ is obtained by taking the $\alpha$-quantile of the bootstrap distribution:

$$\hat{\theta}_\text{PB}[\alpha] = \hat{\theta}^*_\alpha.$$

All implementations of methods that use quantiles, use the median-unbiased version of quantile calculation, recommended in \cite{hyndman1996sample}.

\subsubsection{Standard Bootstrap (\textbf{B-n})}

The standard method (sometimes the normal method), assumes that the bootstrap distribution is normal~\cite[chap.~13]{efron1994introduction}:

$$\hat{\theta}_\text{B-n}[\alpha] = \hat{\theta} + \hat{\sigma} z_\alpha,$$

\noindent where $\hat{\theta}$ is the plug-in estimator of the functional, $\hat{\sigma}$ is the standard deviation estimate from the bootstrap distribution and $z_\alpha$ is the z-score.

\subsubsection{Basic Bootstrap (\textbf{BB})}

In the basic bootstrap~\cite[chap.~13.4]{efron1994introduction}, sometimes called the reverse percentile method, the observed bootstrap distribution $\theta^*$ is replaced with $W^* = \theta^* - \hat{\theta}$. This results in 
$$ \hat{\theta}_\text{BB}[\alpha] = 2\hat{\theta} - \hat{\theta}^*_{1 - \alpha}. $$

Davison and Hinkley \cite{davison1997bootstrap} show that it provides an accurate confidence interval for the sample median, but it can have a substantial coverage error because of errors in quantile calculation of $W^*$. It can also give us invalid parameter values, when there are constraints on $\theta$.

\subsubsection{Smoothed Bootstrap (\textbf{SB})}

The smoothed bootstrap~\citep{davison1997bootstrap} gets its name from smoothing the bootstrap distribution. Smoothing is implemented with a normal kernel centered on 0 and kernel size is determined with

$$ h = 0.9 \min \big( \hat{\sigma}, \frac{\text{IQR}}{1.34} \big),$$

\noindent where $\text{IQR}$ is the inter-quartile range of the bootstrap distribution, respectively. The CI estimate is then obtained by taking the $\alpha$ quantile of the smoothed bootstrap distribution $\Tilde{\theta}^*$:

$$\hat{\theta}_\text{SB}[\alpha] = \Tilde{\theta}^*_\alpha.$$

\subsubsection{Bias Corrected Bootstrap (\textbf{BC})}

The bias corrected bootstrap corrects the bias of the percentile CI~\cite[chap.~14]{efron1994introduction}. The CI estimate is:

\begin{align*}
\hat{\theta}_\text{BC}[\alpha] &= \hat{\theta}^*_{\alpha_{BC}}, \\
\alpha_\text{BC} &= \Phi\big(2\Phi^{-1}(\hat{b}) + z_\alpha \big),
\end{align*}

\noindent where $\Phi$ is the standard normal CDF $\hat{b}$ is the bias, calculated as the percentage of values from the bootstrap distribution that are lower than the value of the functional on the data.

\subsubsection{Bias Corrected and Accelerated Bootstrap (\textbf{BCa})}

The bias corrected and accelerated bootstrap~\cite[chap.~14]{efron1994introduction} further corrects the $BC$ interval by computing acceleration $a$, which accounts for the skewness of the bootstrap distribution:

\begin{align*}
\hat{\theta}_{BCa}[\alpha] &= \hat{\theta}^*_{\alpha_\text{BCa}}, \\
\alpha_\text{BCa} &= \Phi\Big(\Phi^{-1}(b) + \frac{\Phi^{-1}(\hat{b}) + z_\alpha}{1 + \hat{a} (\Phi^{-1}(\hat{b}) + z_\alpha)} \Big),
\end{align*}

\noindent where $\hat{a}$ is the leave-one-out jackknife approximation of the acceleration constant.

\subsubsection{Studentized Bootstrap (\textbf{B-t})}

The studentized bootstrap~\cite[chap.~14]{efron1994introduction}, also known as bootstrap-t, generalizes the Student's t method, using the distribution of $T = (\hat{\theta} - \theta) /\hat{\sigma}$ to estimate the CI:

$$\hat{\theta}_\text{B-t}[\alpha] = \hat{\theta} - \hat{\sigma} T_{1-\alpha}.$$

But since the distribution of $T$ is not known, its percentiles have to be approximated from the bootstrap distribution. That is done by defining $T^* = (\hat{\theta}^* - \hat\theta) / \hat{\sigma}^*$, where $\hat{\sigma}^*$ is obtained by doing another inner bootstrap sampling on each of the outer samples.

\subsubsection{Double Bootstrap (\textbf{DB})}

The double bootstrap~\cite[chap.~3.11]{hall2013bootstrap} corrects bias with another level of inner of bootstraps. The bootstrap procedure is repeated on each of the bootstrap samples to calculate the the percentage of times that the inner bootstrap functional is smaller than on the original sample. A limit is required such that $P \{\hat{\theta} \in (-\infty, \hat{\theta}_{double}[\alpha])\} = \alpha$, which is why the $\alpha$-th quantile of biases $\hat{b}^*$ is selected for the adjusted level $\alpha_\text{DB}$. This leads to

\begin{align*}
\hat{\theta}_\text{DB}[\alpha] &= \hat{\theta}^*_{\alpha_{double}}, \\ 
\alpha_\text{DB} &= \hat{b}^*_\alpha.
\end{align*}

Note that percentile bootstrap is used on the inner and outer bootstrap, but the double (iterated) bootstrap allows for any bootstrap method.
    
\subsection{Measuring the quality of confidence intervals}\label{sec:measuring}

Most related work measures only coverage, with only a few studies measuring interval length \citep{carpenter2000bootstrap,DelosReyes2023,efron2003second,flowers2018comparison,Ialongo2019,owen1992empirical,puth2015} or comparing CIs with exact intervals \citep{diciccio1996bootstrap,diciccio1988review,efron2003second}.

This study also focuses on coverage, but also measures and reports results for \textit{interval length} for two-sided CIs and the \textit{absolute distance from exact intervals} for one-sided intervals. The exact interval for endpoint $\alpha$ and parameter $\theta$ is defined as $\hat{\theta}_{exact}[\alpha] = \hat{\theta} - \hat{\sigma}K^{-1}(1 - \alpha)$, where $K$ is the cumulative distribution function of $\theta$ \citep{DiCiccioRejoinder1996}. Note that $\hat{\sigma}K^{-1}(1-\alpha)$ is approximated with 100000 samples.

Note that two-sided coverage can be misleading regarding a method's coverage because good two-sided coverage can be, and in practice often is, a result of substantial, but opposite errors in the two one-sided intervals (see \cite{DiCiccioRejoinder1996} for an example). That is, while two-sided error can be studied from one-sided CIs, the converse is not true.

The practical meaning of coverage error depends on nominal coverage and is not symmetric. For example, 51\% coverage at 50\% nominal coverage is not the same as 96\% coverage at 95\% nominal coverage. And 85\% coverage is not the same as 95\% coverage at 90\% nominal coverage. To aggregate results and for a threshold-based criterion that can be applied to all confidence levels, a novel criterion is proposed, based on the Kullback-Leibler divergence. That is, it is based on measuring information loss if nominal coverage $\pi$ is assumed when actual coverage is $p$:

$$\text{KL}(p,\pi) = p \log_2(\frac{p}{\pi}) + (1-p) \log_2(\frac{1 - p}{1 - \pi}).$$  

For a threshold-based criterion of what is considered \emph{good enough} Bradley's criterion $|p - \pi| < \frac{\min(\pi, 1 - \pi)}{k}$ is modified, where $\pi$ is the nominal coverage \citep{bradley1978robustness}. Common choices for $k$ are 10 (stringent), 4 (intermediate), 2 (liberal), and 0.75 (very liberal). Note that the intermediate and very liberal were introduced by \cite{robey1992type}. In related work on the bootstrap, only \cite{DelosReyes2023} uses a \emph{good enough} criterion - that actual coverage should lie between 92.5\% and 97.5\% when nominal coverage is 95\% (this is based on the work of \cite{bradley1978robustness}).

At $95\%$ nominal coverage, the KL divergences for the Bradley lower bounds $94.5\%$ ($k = 10$), $93.8\%$ ($k = 4$), $92.5\%$ ($k = 2$), and $88.3\%$ ($k = 0.75$) are approximately $0.0004$, $0.0020$, $0.0083$, and $0.0503$. The value $\text{KL}(0.945, 0.95)$ is adopted as the stringent criterion and a factor of 5 as an \emph{order of magnitude} worse/better performance. For nominal coverage 95\% this leads to criteria very similar to Bradley's: $(93.8, 96.2)$ vs $(93.9, 96.1)$ (intermediate), $(92.5, 97.5)$ vs $(92.4, 97.3)$ (liberal), and $(88.3, 101.7)$ vs $(88.6, 99.4)$ (very liberal). However, for nominal coverage further away from 95\%, the proposed approach gives produces sensible criteria and does not produce endpoints outside of the unit interval. 

\section{Results} \label{ch:results}

\begin{figure}[tp]
    \centering
\includegraphics[width=0.99\linewidth]{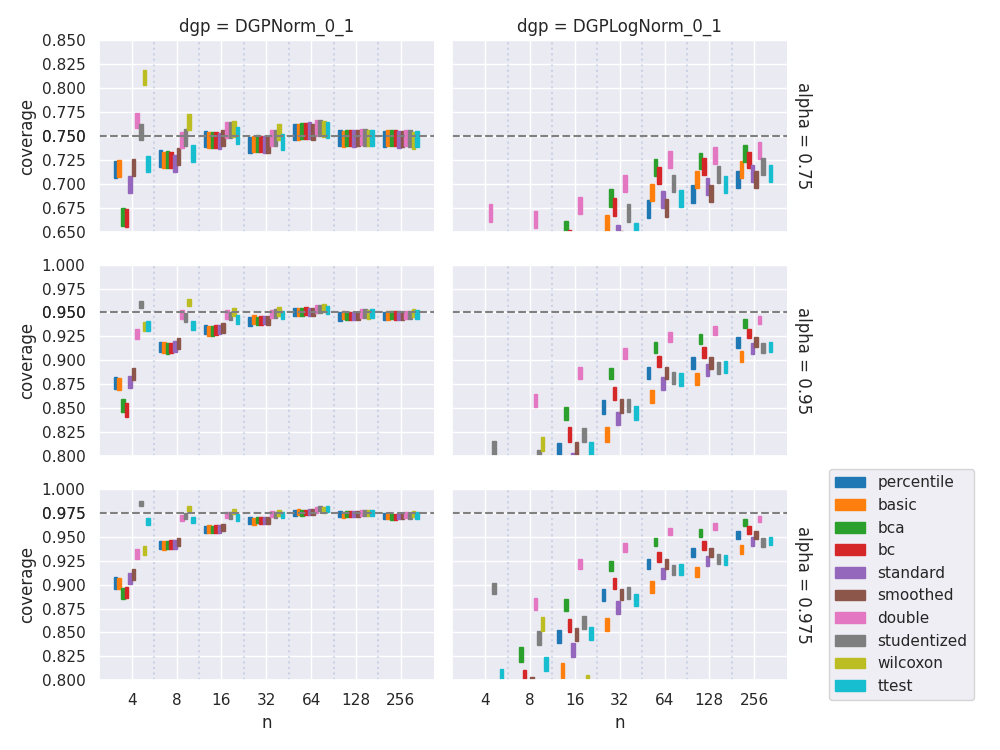}
    \caption{A visualization of the visualization tool for browsing the results of the simulation study. The user can filter results by method and facet based on the dimensions of the experiment (functional, confidence level, distribution). }
    \label{fig:app}
\end{figure}

The simulation study consists of 1386 combinations of sample size, endpoint, and functional, and it is infeasible to list all of the results. The results first focus on identifying if a bootstrap method is a viable one-size-fits-all approach. Inevitably, some details that are relevant to the reader might be left out. A visualization tool is available to browse all of the results of the experiments {\small\url{zrimseku.github.io/bootstrap-simulation/}} (see Figure \ref{fig:app}).

The raw results (aggregated over 10000 replications) and the source code for the tool, simulation framework, pre-processing, and analysis can be found here: {\small\url{github.com/zrimseku/bootstrap-simulation}}. The source code can be used to generate the full non-aggregated results. The library with all the methods can be found here: {\small\url{github.com/zrimseku/bootstrap-ci}}.

\subsection{When methods fail to produce a CI}

There are only a few cases where a method fails to produce a CI. All bootstrap methods fail to produce a CI for Pearson correlation for $n = 4$, due to division by zero variance. It can also happen for $n = 8$, but rarely. BC and BCa do not produce a CI for the $5^{th}$ percentile for $n \leq 8$. For the $95^{th}$ percentile, B-t does not produce an interval in most cases when $n = 4$ for the Laplace distribution. m-j is unable to produce CIs for small sample sizes and extreme percentiles ($5^{th}$ percentile for $n \leq 16$ and the $95^{th}$ percentile for $n \leq 8$). Method q-nonpar fails to produce the $95^{th}$ percentile for $n = 4$ and $\alpha \leq 0.75$, for $n \in \{8, 16\}$ and $\alpha \leq 0.25$, for $n = 32$ and $\alpha \leq 0.05$ and for $n = 64$ and $alpha= 0.025$. When predicting CIs for the median, it fails at $n=4$ and $\alpha \leq 0.05$. And, although wilcoxon returns CIs for asymmetric distributions, they are not useful CIs for location. Also note that all cases for Bernoulli, where all $n$ had the same value, are removed, because most methods fail if there is no variability in the data.

\subsection{Coverage of bootstrap methods}\label{sec:coverage}

Table \ref{tbl:compare_kl_one_sided} shows a comparison of bootstrap methods in mean $\text{KL}$. As expected, coverage improves with sample size and the two extreme percentiles and standard deviation are the most difficult functionals. For Pearson correlation, mean, and median, DB is best. For the percentiles, B-n is best. And for standard deviation, B-t is best. B-n and B-t perform the best overall. Because coverage gets wors with smaller sample size, the overall results are biased towards methods that perform well on small $n$. DB and BCa, which are expected to perform best, perform relatively poorly for small $n$, but are best and second best for sample sizes $n \geq 64$.

\begin{table}[tbp]
\tbl{\textbf{Mean $\text{KL}$ coverage performance of bootstrap methods for one-sided CIs.} The \textit{all} column is across all combinations, while the remaining results are grouped by sample size or statistical functional. The best performing method for each column is \underline{underlined}.}{
\begin{tabular}{lrrrrrrrrrrrrrr}
  \toprule
 & all & 4 & 8 & 16 & 32 & 64 & 128 & 256 & corr & mean & Q$_{0.5}$ & Q$_{0.05}$ & Q$_{0.95}$ & std \\ 
  \midrule
B-n & $\underline{0.078}$ & $0.317$ & $\underline{0.174}$ & $\underline{0.087}$ & $\underline{0.037}$ & $0.021$ & $0.014$ & $0.009$ & $0.006$ & $0.026$ & $0.004$ & $\underline{0.024}$ & $\underline{0.145}$ & $0.208$ \\ 
  B-t & $0.084$ & $\underline{0.196}$ & $0.269$ & $0.107$ & $0.049$ & $0.023$ & $0.019$ & $0.014$ & $0.008$ & $\underline{0.013}$ & $0.012$ & $0.078$ & $0.264$ & $\underline{0.084}$ \\ 
  BB & $0.112$ & $0.249$ & $0.189$ & $0.218$ & $0.101$ & $0.046$ & $0.033$ & $0.022$ & $0.017$ & $0.031$ & $0.045$ & $0.159$ & $0.234$ & $0.145$ \\ 
  SB & $0.118$ & $0.452$ & $0.267$ & $0.152$ & $0.060$ & $0.027$ & $0.017$ & $0.011$ & $0.002$ & $0.022$ & $0.003$ & $0.064$ & $0.230$ & $0.308$ \\ 
  DB & $0.134$ & $0.536$ & $0.340$ & $0.193$ & $0.057$ & $\underline{0.008}$ & $\underline{0.005}$ & $\underline{0.003}$ & $\underline{0.000}$ & $\underline{0.013}$ & $\underline{0.002}$ & $0.125$ & $0.401$ & $0.188$ \\ 
  PB & $0.157$ & $0.610$ & $0.364$ & $0.219$ & $0.073$ & $0.028$ & $0.018$ & $0.011$ & $0.002$ & $0.025$ & $0.003$ & $0.109$ & $0.377$ & $0.330$ \\ 
  BC & $0.161$ & $0.628$ & $0.359$ & $0.228$ & $0.092$ & $0.022$ & $0.013$ & $0.008$ & $0.001$ & $0.049$ & $0.017$ & $0.169$ & $0.387$ & $0.246$ \\ 
  BCa & $0.161$ & $0.632$ & $0.359$ & $0.236$ & $0.092$ & $0.018$ & $0.010$ & $0.005$ & $0.001$ & $0.058$ & $0.017$ & $0.201$ & $0.379$ & $0.218$ \\  
\bottomrule
\end{tabular}}
\label{tbl:compare_kl_one_sided}
\end{table}

\begin{table}[tbp]
\tbl{\textbf{Threshold-based coverage performance of bootstrapping on one-sided CIs.} The table shows the percentage of experiments where a method does not meet the liberal criterion $25 \times \text{KL}(0.945, 0.95)$. The \textit{all} column is across all experiments, while the remaining results are grouped by sample size or statistical functional. The best performing method for each column is \underline{underlined}.}{
\begin{tabular}{lrrrrrrrrrrrrrr}
  \toprule
 & all & 4 & 8 & 16 & 32 & 64 & 128 & 256 & corr & mean & Q$_{0.5}$ & Q$_{0.05}$ & Q$_{0.95}$ & std \\ 
  \midrule
DB & $\underline{0.30}$ & $0.67$ & $\underline{0.49}$ & $\underline{0.46}$ & $0.44$ & $\underline{0.12}$ & $\underline{0.07}$ & $\underline{0.03}$ & $\underline{0.00}$ & $\underline{0.18}$ & $\underline{0.05}$ & $0.46$ & $0.50$ & $\underline{0.42}$ \\ 
  B-n & $0.40$ & $0.78$ & $0.65$ & $0.56$ & $\underline{0.36}$ & $0.28$ & $0.20$ & $0.13$ & $0.20$ & $0.34$ & $0.11$ & $\underline{0.28}$ & $0.46$ & $0.80$ \\ 
  BCa & $0.41$ & $0.89$ & $0.70$ & $0.62$ & $0.41$ & $0.23$ & $0.21$ & $0.06$ & $\underline{0.00}$ & $0.32$ & $0.33$ & $0.68$ & $\underline{0.41}$ & $0.47$ \\ 
  SB & $0.45$ & $0.83$ & $0.63$ & $0.57$ & $0.49$ & $0.40$ & $0.28$ & $0.14$ & $0.07$ & $0.28$ & $0.12$ & $0.49$ & $0.57$ & $0.94$ \\ 
  BC & $0.46$ & $0.89$ & $0.72$ & $0.68$ & $0.50$ & $0.34$ & $0.24$ & $0.11$ & $\underline{0.00}$ & $0.31$ & $0.33$ & $0.57$ & $0.73$ & $0.54$ \\ 
  PB & $0.47$ & $0.82$ & $0.66$ & $0.63$ & $0.51$ & $0.39$ & $0.28$ & $0.16$ & $0.07$ & $0.32$ & $0.10$ & $0.49$ & $0.59$ & $0.94$ \\ 
  B-t & $0.47$ & $\underline{0.54}$ & $0.62$ & $0.55$ & $0.51$ & $0.43$ & $0.41$ & $0.32$ & $0.23$ & $0.30$ & $0.35$ & $0.59$ & $0.67$ & $0.57$ \\ 
  BB & $0.57$ & $0.80$ & $0.75$ & $0.64$ & $0.60$ & $0.50$ & $0.46$ & $0.41$ & $0.27$ & $0.37$ & $0.59$ & $0.76$ & $0.81$ & $0.50$ \\   
\bottomrule 
\end{tabular}}
\label{tbl:compare_coverage_one_sided}
\end{table}

\begin{table}[tbp]
\tbl{\textbf{Threshold-based coverage performance of bootstrapping on two-sided CIs.} The table shows the percentage of experiments where a method does not meet the liberal criterion $25 \times \text{KL}(0.945, 0.95)$. The \textit{all} column is across all experiments, while the remaining results are grouped by sample size or statistical functional. The best performing method for each column is \underline{underlined}.}{
\begin{tabular}{lrrrrrrrrrrrrrr}
\toprule
 & all & 4 & 8 & 16 & 32 & 64 & 128 & 256 & corr & mean & Q$_{0.5}$ & Q$_{0.05}$ & Q$_{0.95}$ & std \\ 
  \midrule
DB & $\underline{0.33}$ & $0.82$ & $\underline{0.53}$ & $\underline{0.58}$ & $\underline{0.53}$ & $\underline{0.11}$ & $\underline{0.03}$ & $\underline{0.03}$ & $\underline{0.00}$ & $0.29$ & $\underline{0.07}$ & $0.43$ & $\underline{0.40}$ & $\underline{0.57}$ \\ 
  SB & $0.44$ & $1.00$ & $0.68$ & $0.71$ & $0.62$ & $0.23$ & $0.12$ & $0.08$ & $0.10$ & $0.44$ & $0.14$ & $0.43$ & $0.42$ & $0.82$ \\ 
  B-n & $0.46$ & $0.97$ & $0.82$ & $0.76$ & $0.56$ & $0.29$ & $0.12$ & $0.08$ & $0.40$ & $0.46$ & $0.23$ & $\underline{0.35}$ & $0.43$ & $0.80$ \\ 
  BCa & $0.47$ & $1.00$ & $0.89$ & $0.68$ & $0.59$ & $0.33$ & $0.11$ & $0.06$ & $\underline{0.00}$ & $0.42$ & $0.25$ & $0.62$ & $\underline{0.40}$ & $0.74$ \\ 
  PB & $0.47$ & $1.00$ & $0.76$ & $0.76$ & $0.62$ & $0.30$ & $0.12$ & $0.08$ & $0.20$ & $0.44$ & $0.19$ & $0.47$ & $0.47$ & $0.82$ \\ 
  BC & $0.48$ & $1.00$ & $0.92$ & $0.73$ & $0.58$ & $0.36$ & $0.09$ & $0.06$ & $\underline{0.00}$ & $0.43$ & $0.25$ & $0.57$ & $0.45$ & $0.77$ \\ 
  B-t & $0.66$ & $\underline{0.53}$ & $0.74$ & $0.80$ & $0.74$ & $0.62$ & $0.58$ & $0.56$ & $0.40$ & $\underline{0.21}$ & $0.76$ & $1.00$ & $1.00$ & $0.65$ \\ 
  BB & $0.80$ & $1.00$ & $1.00$ & $0.92$ & $0.83$ & $0.76$ & $0.64$ & $0.59$ & $0.60$ & $0.47$ & $0.95$ & $1.00$ & $1.00$ & $0.79$ \\ 
 \bottomrule
\end{tabular}}
\label{tbl:compare_coverage_two_sided}
\end{table}

Table \ref{tbl:compare_coverage_one_sided} shows how often methods fail to meet the stringent criterion. DB outperforms other methods overall and on all sample sizes except $n = 4$. It is the best method or relatively close to the best on all functionals, except on one of the two extreme percentiles. BCa is second-best overall. The results so far suggest that overall DB is best, but can have very poor coverage in some cases, especially for the two extreme percentiles at small sample sizes. While a more liberal criterion will result in fewer failures for all methods, the ordering does not change.

Four of the six confidence levels in the experiments can be used to derive results for the two most common two-sided CIs (95\%, 90\%). Table \ref{tbl:compare_coverage_two_sided} shows how often the methods fail to meet the liberal criterion for these two-sided intervals. These and results for other thresholds and $\text{KL}$ are similar for one-sided and two-sided intervals. Most bootstrap methods meet the liberal criterion in almost all cases when $n \geq 128$ ($n \geq 64$ for DB), but even at $n = 256$ there are still two cases where even DB does not (standard deviation for log-normal). Excluding these, DB meets the very liberal criterion for all experiments for $n \geq 64$.

\subsection{Coverage comparison with baseline methods} \label{sec:comparison}

All of the results in this section are for one-sided CIs. Results for two-sided CIs are similar. Starting with the premise that when limited to a single method, DB is the best choice, this section takes a closer look at where DB is clearly outperformed by another method. The criterion used is that method A outperforms method B if B does not meet the liberal criterion and A is at least an order of magnitude better than method B. If a method is not able to produce CIs, it is outperformed by any method that can. 

\begin{table}[tp]
\tbl{\textbf{Cases where DB and baseline outperform each other.} The number in the parentheses is the number of combinations where a method outperforms. There are a total of 36 combinations for every pair of $n$ and functional, except Pearson correlation, where there are 6, and the mean, where there are 48.}{
\begin{tabular}{rrrrr}
  \toprule
 n & functional & baseline $\gg$ DB & DB $ \gg$ baseline  \\ 
  \midrule
  4 & corr & fisher (6) &  \\ 
    4 & mean & a-c (2); c-p (1); t-test (11); wilcoxon (1) & DB (2) \\ 
    4 & median & m-j (5); q-par (3) &  \\ 
    4 & Q(0.05) & q-nonpar (12); q-par (22) &  \\ 
    4 & Q(0.95) & q-par (21) & DB (4) \\ 
    4 & std & chi-sq (14) & DB (5) \\ 
    8 & corr & fisher (6) &  \\ 
    8 & mean & a-c (7); c-p (5); t-test (10); wilcoxon (3) & DB (3) \\ 
    8 & Q(0.05) & q-nonpar (12); q-par (22) & DB (3) \\ 
    8 & Q(0.95) & q-nonpar (12); q-par (21) &  \\ 
    8 & std & chi-sq (7) & DB (9) \\ 
   16 & mean & a-c (4); c-p (3); t-test (4) & DB (9) \\ 
   16 & Q(0.05) & q-nonpar (17); q-par (15) &  \\ 
   16 & Q(0.95) & m-j (10); q-par (11) & DB (4) \\ 
   16 & std & chi-sq (2) & DB (9) \\ 
   32 & mean & a-c (1); c-p (1); t-test (3) & DB (8) \\ 
   32 & Q(0.05) & m-j (6); q-nonpar (17); q-par (5) &  \\ 
   32 & Q(0.95) & m-j (17); q-par (6) &  \\ 
   32 & std &  & DB (12) \\ 
   64 & mean & a-c (1); c-p (1) & DB (7) \\ 
   64 & Q(0.05) & m-j (1); q-nonpar (3) &  \\ 
   64 & Q(0.95) &  & DB (5) \\ 
   64 & std & chi-sq (1) & DB (18) \\ 
  128 & mean & a-c (2); c-p (1); t-test (1) & DB (6) \\ 
  128 & Q(0.05) & m-j (2) & DB (1) \\ 
  128 & std &  & DB (24) \\ 
  256 & mean &  & DB (4) \\ 
  256 & std &  & DB (27) \\ 
 \bottomrule
\end{tabular}}
\label{tbl:baseline_double}
\end{table}

Table \ref{tbl:baseline_double} adds detail to the results from Section \ref{sec:coverage}. For Pearson correlation DB does not produce CIs for $n = 4$ and performs poorly for $n = 8$. It performs poorly on the two extreme percentiles for $n \leq 32$ and in some cases on the other functionals, mostly for $n \leq 8$. For Q(0.05) and Q(0.95), a better choice is q-par for the smaller sample sizes and m-j for sample size $n = 32$. For the other functionals and $n \geq 8$ ($n \geq 16$ for Pearson correlation) there is no clear advantage of using baseline methods. Note that for Q(0.05) and Q(0.95) and $n = 16, 32$ B-n is as good as or better than baseline. Table \ref{tbl:baseline_standard} shows the advantage of using B-n instead of DB for Q(0.05) and Q(0.95) for $n \leq 32$.

\begin{table}[tp]
\tbl{\textbf{Cases where B-n and baseline outperform each other.} Results are for the two extreme percentiles and $n \leq 32$. The number in the parentheses is the number of combinations where a method outperforms. There are a total of 36 combinations for every pair of $n$ and functional.}{
\begin{tabular}{rrrrr}
  \toprule
 n & functional & baseline $\gg$ B-n & B-n $\gg$ baseline \\ 
  \midrule
    4 & Q(0.05) & q-nonpar (11); q-par (14) &  \\ 
    4 & Q(0.95) & q-par (14) &  \\ 
    8 & Q(0.05) & q-par (11) & B-n (1) \\ 
    8 & Q(0.95) & q-nonpar (4); q-par (10) & B-n (3) \\ 
   16 & Q(0.05) & q-nonpar (5); q-par (6) & B-n (5) \\ 
   16 & Q(0.95) & m-j (1); q-par (6) &  \\ 
   32 & Q(0.05) &  & B-n (4) \\ 
   32 & Q(0.95) & m-j (2); q-par (1) & B-n (6) \\ 
\bottomrule
\end{tabular}}
\label{tbl:baseline_standard}
\end{table}

Note that BCa behaves similarly to DB, but DB is the better choice. For example, DB is at least an order of magnitude better than BCa in 17\% of all cases while BCa is better than DB in 4\% of the cases.

\subsection{Absolute distance from exact CIs} \label{sec:length}

Coverage by itself is not sufficient, as CI endpoints should also match the exact endpoints. An extreme example would be to generate a very large endpoint (wide two-sided interval) with probability $\alpha$ and a very small endpoint (narrow two-sided interval) with probability $\alpha$. This would result in nominal coverage but useless CIs.

\begin{figure}[tp]
    \centering
\includegraphics[width=0.75\linewidth]{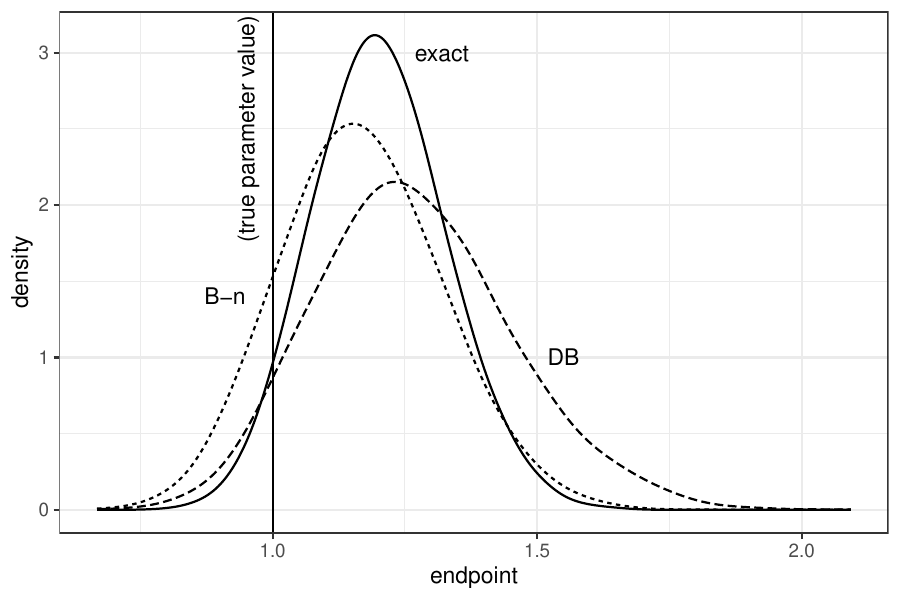}
    \caption{Density estimates for endpoints for $n = 32$, $\pi = 0.95$, standard normal DGP, and standard deviation. Estimated coverage of DB and B-t is 93.7\% and 86.3\%, respectively. Absolute distance from exact is 0.084 and 0.045, respectively. Note that the endpoints produced by methods strongly correlate with exact endpoints ($\rho \approx 0.9$).}
    \label{fig:example}
\end{figure}

Table \ref{tbl:comparison_dist_to_exact} shows that B-n has lowest distance from exact, while DB and BCa perform relatively poorly across all groups. This result cannot be interpreted in isolation, because there is typically a trade-off between coverage and distance (see Figure \ref{fig:example} for an illustrative example).

\begin{table}[tbp]
\tbl{\textbf{Mean absolute distance from exact for one-sided CIs.} For each combination, the value is normalized with two standard deviations of exact intervals. The \textit{all} column is across all experiments, while the remaining results are grouped by sample size or statistical functional. The best performing method for each column is \underline{underlined}.}{
\begin{tabular}{lrrrrrrrrrrrrrr}
  \toprule
 & all & 4 & 8 & 16 & 32 & 64 & 128 & 256 & corr & mean & Q$_{0.5}$ & Q$_{0.05}$ & Q$_{0.95}$ & std \\ 
  \midrule
B-n & $\underline{0.240}$ & $0.423$ & $\underline{0.333}$ & $\underline{0.269}$ & $\underline{0.221}$ & $\underline{0.180}$ & $\underline{0.145}$ & $\underline{0.118}$ & $\underline{0.095}$ & $\underline{0.146}$ & $\underline{0.201}$ & $\underline{0.329}$ & $\underline{0.330}$ & $0.244$ \\ 
  BB & $0.264$ & $\underline{0.395}$ & $0.346$ & $0.317$ & $0.261$ & $0.206$ & $0.178$ & $0.153$ & $\underline{0.095}$ & $\underline{0.146}$ & $0.262$ & $0.362$ & $0.383$ & $\underline{0.227}$ \\ 
  SB & $0.318$ & $0.544$ & $0.439$ & $0.352$ & $0.295$ & $0.243$ & $0.199$ & $0.165$ & $0.125$ & $0.161$ & $0.276$ & $0.479$ & $0.441$ & $0.308$ \\ 
  BC & $0.332$ & $0.620$ & $0.461$ & $0.382$ & $0.316$ & $0.253$ & $0.208$ & $0.173$ & $0.136$ & $0.242$ & $0.286$ & $0.416$ & $0.536$ & $0.258$ \\ 
  BCa & $0.340$ & $0.600$ & $0.458$ & $0.392$ & $0.329$ & $0.275$ & $0.222$ & $0.183$ & $0.136$ & $0.268$ & $0.286$ & $0.432$ & $0.521$ & $0.268$ \\ 
  PB & $0.350$ & $0.642$ & $0.499$ & $0.386$ & $0.313$ & $0.249$ & $0.205$ & $0.170$ & $0.124$ & $0.166$ & $0.295$ & $0.554$ & $0.502$ & $0.320$ \\ 
  DB & $0.360$ & $0.597$ & $0.506$ & $0.394$ & $0.329$ & $0.286$ & $0.234$ & $0.188$ & $0.159$ & $0.284$ & $0.299$ & $0.488$ & $0.501$ & $0.279$ \\ 
  B-t & $0.966$ & $4.376$ & $0.673$ & $0.668$ & $0.524$ & $0.300$ & $0.231$ & $0.188$ & $0.116$ & $0.624$ & $1.751$ & $0.912$ & $1.127$ & $0.636$ \\ 
 \bottomrule
\end{tabular}}
\label{tbl:comparison_dist_to_exact}
\end{table}

Table \ref{tbl:comparison_dist_to_exact_outperform} shows where a baseline method has better distance from exact than DB, but only for cases where its coverage is not an order of magnitude worse. B-n is included as the bootstrap method that performs best in distance from exact. Results are similar to those in Table \ref{tbl:baseline_double} - baseline methods and B-n outperform DB for the two extreme percentiles. That is, in most cases, lower distance from exact comes at the expense of worse coverage.

\begin{table}[tp]
\tbl{\textbf{Cases where baseline (with B-n included) outperforms DB in absolute distance from exact intervals.} The number in the parentheses is the number of combinations where a method outperforms. Note that there are a total of 36 combinations for every pair of $n$ and functional, except Pearson correlation, where there are 6, and the mean, where there are 48.}{
\begin{tabular}{rrrrr}
  \toprule
 n & functional & baseline and B-n $\gg$ DB  \\ 
  \midrule
  4 & corr & fisher (6)   \\ 
    4 & mean & a-c (2); B-n (7); c-p (1); t-test (3); wilcoxon (1) & \\ 
    4 & median & B-n (2); m-j (2); q-par (2)   \\ 
    4 & Q(0.05) & B-n (19); q-nonpar (3); q-par (17)   \\ 
    4 & Q(0.95) & B-n (17); q-par (15)   \\ 
    4 & std & B-n (6) & \\ 
    8 & corr & fisher (6)   \\ 
    8 & mean & a-c (2); B-n (6); c-p (1); t-test (8); wilcoxon (2)   \\ 
    8 & Q(0.05) & B-n (21); q-nonpar (6); q-par (18)   \\ 
    8 & Q(0.95) & B-n (19); q-nonpar (7); q-par (19)   \\ 
   16 & corr & fisher (1)   \\ 
   16 & mean & B-n (4); t-test (3)   \\ 
   16 & Q(0.05) & B-n (11); q-nonpar (3); q-par (6)   \\ 
   16 & Q(0.95) & B-n (8); m-j (6); q-par (6)   \\ 
   32 & mean & a-c (1); B-n (3); c-p (1); t-test (5)   \\ 
   32 & Q(0.05) & B-n (8); m-j (5); q-par (9)   \\ 
   32 & Q(0.95) & B-n (8); m-j (4); q-par (11)   \\ 
   32 & std & chi-sq (1)   \\ 
   64 & mean & B-n (1); t-test (1)   \\ 
   64 & Q(0.05) & B-n (3); m-j (1); q-par (1)   \\ 
   64 & Q(0.95) & B-n (1); m-j (6); q-par (1)   \\ 
   64 & std & chi-sq (2)   \\ 
  128 & mean & B-n (1); t-test (1)   \\ 
  128 & Q(0.05) & B-n (1); m-j (1); q-nonpar (1)   \\ 
  128 & Q(0.95) & B-n (2); m-j (3)   \\ 
 \bottomrule
\end{tabular}}
\label{tbl:comparison_dist_to_exact_outperform}
\end{table}

\section{Conclusion}

This paper is the most comprehensive review of empirical results for bootstrap methods to date, along with an extensive empirical comparison. The simulation study encompasses not only the most widely used non-parametric bootstrap methods but also the most common techniques for quantifying uncertainty in general. Furthermore, a novel criterion for assessing confidence interval quality is introduced, which improves on existing evaluation approaches.

DB is identified as the overall best method, contrary to some recommendations of BCa for general cases \citep{carpenter2000bootstrap}, but in line with other recommendations \citep{Hall1996comment,Lee1996} and the few empirical studies that included DB. These results are also in line with related work: PB performs relatively poorly. For small $n$ B-t performs best for the mean and BCa best for Pearson correlation, although DB performs similarly. B-t relies on an estimate of variance and can produce very long CIs. Bootstrapping can perform poorly compared do chi-squared CIs on variance for small $n$ and normal distribution, however, as demonstrated, chi-squared performs poorly on non-normal distributions and DB is comparable or better for $n \geq 32$.

DB has two weaknesses. It can perform relatively poorly when $n = 4, 8$ and on extreme percentiles for $n \leq 32$. The latter can be mitigated by using B-n but also raises the question if DB can be modified to deal with these cases, while still preserving most of its simplicity. This leaves us with $n = 4, 8$, where it has to be acknowledged that a non-parametric approach is often worse than a parametric approach even when the assumptions of the parametric method are violated. It was not investigated to what extent bootstrap diagnostics (see \cite{canty2006bootstrap} and \cite[Ch.3.10]{davison1997bootstrap}) could further mitigate these issues.

The dimensions of the experiment can be improved, as it only includes one DGP for Pearson correlation, which is not nearly as comprehensive as the recent study by \cite{DelosReyes2023}, who did not include DB. Other commonly used functionals could also be included, such as regression model coefficients, non-parametric correlation, and distances between distributions. Furthermore, the study focuses on independent data. Related work on hierarchical, temporal, spatial, and other dependencies is sparse, but mostly in favor of bootstrapping. And in more complex problems bootstrapping is often the only viable approach.

\subsection{Practical implications}

One general implication of these results for practitioners is that DB is superior to BCa. Considering that the DB is also conceptually simpler and easier to implement than the BCa, it should be a standard feature in every bootstrap library. This is currently not the case. For example, SPSS and popular bootstrap libraries from Python (scipy.stats.bootstrap, bootstrapping) and R (boot) do not implement any form of iterated bootstrap. While the boot package does implement the studentized bootstrap, which involves inner bootstrap sampling, this method performed poorly in our simulation study.

The underutilization of the double bootstrap could be attributed to the misconception that BCa is superior or to the DB being more computationally intensive. Due to its quadratic time complexity, the double bootstrap will not scale to large sample sizes as well as BCa. However, with advances in computation, this is no longer a significant issue. It can also be argued that once a sample size is large enough to make the double bootstrap prohibitive, the choice of method becomes moot, as even the percentile bootstrap will likely perform well enough. Additionally, the double bootstrap is easy to parallelize, and efficient approximations have been developed, such as \cite{Chang2015DoubleBootstrap,davidson2020fast}.

In critical applications of statistics, there is no substitute for thoroughly understanding the statistical task at hand, carefully choosing the most appropriate method, and applying it correctly. However, if a single method had to be recommended, the double bootstrap appears to be the best choice. The two scenarios where particular care should be taken are with very small sample sizes and when estimating extreme quantiles, both of which are arguably less common in practice.

The double bootstrap can also play an important role in the teaching of applied statistics. In particular, it can be a valuable tool in a bootstrap-centric introduction to applied statistics for a broad audience of individuals who will use statistics but will only receive one or two courses of formal statistical training. For example, in most empirical sciences.

A basic understanding of standard statistical functionals is required, regardless of what method is used to quantify uncertainty. For example, the mean (average) has to be understood, before bootstrapping, t-distribution-based, or any other approach to constructing CIs can be used. Once functionals are understood, percentile bootstrap is conceptually very simple, can be implemented in a few lines of code, and then applied in the same way to any functional. This is an order of magnitude less complex and time-consuming than teaching a different approach for every functional.

The main disadvantage of the percentile bootstrap is its suboptimal performance in practice. Consequently, a student transitioning into a practitioner must either switch to more complicated bootstrap approaches, such as BCa, or apply parametric approaches. The double bootstrap alleviates this issue, similar to the percentile bootstrap. With just a few more lines of code, we can achieve excellent performance, often surpassing methods that are more complex to understand and implement.

This bootstrap-centric approach is less pedagogically complex than learning several different approaches, and it is computationally not an issue on modern hardware. Lack of software support might be an issue, but a minor one, because percentile and double bootstrap are easy to implement. We agree with \cite{Hesterberg2015} that the main reasons that such a shift has not occurred is the lack of teaching materials and legitimate concerns for backwards compatibility. That is, the concern with not teaching the students the methods that are currently dominant statistical practice. This study contributes to alleviating those concerns, showing that bootstrapping, in particular double bootstrap, is overall at least as good as, if not superior to, standard parametric approaches. However, the logical next step is to develop and test an introductory course in applied statistics that relies solely on the bootstrap.

\section*{Acknowledgment}

We thank Gregor Sočan for his helpful comments.

\section*{Funding}

This work was supported by the Slovenian Research and Innovation Agency under Grants P2-0442 and J5-60084. 

\section*{Data Availability Statement}

The authors confirm that the data supporting the findings of this study are available within the article and its supplementary materials.

\bibliographystyle{tfq}
\bibliography{main}

\end{document}